\documentclass[a4paper,11pt]{article}
\pdfoutput=1 

\usepackage{jcappub} 
\usepackage{xspace}
\usepackage{mathtools}
\usepackage[T1]{fontenc} 
\newcommand{\Xm}{\ensuremath{X_{\rm max}}\xspace}
\usepackage{multirow}
\usepackage[switch, modulo]{lineno}
\usepackage{tabularx, booktabs}
\newcolumntype{Y}{>{\centering\arraybackslash}X}
\usepackage{floatrow}
\newfloatcommand{capbtabbox}{table}[][\FBwidth]
\usepackage{graphicx}
\usepackage{capt-of}
\title{\bf Search for photons with energies above 10$^{18}$~eV using the hybrid detector of the Pierre Auger Observatory}

\collaboration{The Pierre Auger Collaboration}

\author[63]{A.~Aab,}
\author[70]{P.~Abreu,}
\author[48,47]{M.~Aglietta,}
\author[29]{I.~Al Samarai,}
\author[16]{I.F.M.~Albuquerque,}
\author[1]{I.~Allekotte,}
\author[8,11]{A.~Almela,}
\author[62]{J.~Alvarez Castillo,}
\author[79]{J.~Alvarez-Mu\~niz,}
\author[38]{G.A.~Anastasi,}
% ...

\affiliation[1]{Centro At\'omico Bariloche and Instituto Balseiro (CNEA-UNCuyo-CONICET), Argentina}
% ...
\affiliation[8]{Instituto de Tecnolog\'\i{}as en Detecci\'on y Astropart\'\i{}culas (CNEA, CONICET, UNSAM), Centro At\'omico Constituyentes, Comisi\'on Nacional de Energ\'\i{}a At\'omica, Argentina}
% ...
\affiliation[11]{Universidad Tecnol\'ogica Nacional -- Facultad Regional Buenos Aires, Argentina}
% ...
\affiliation[16]{Universidade de S\~ao Paulo, Inst.\ de F\'\i{}sica, S\~ao Paulo, Brazil}
% ...
\affiliation[29]{Laboratoire de Physique Nucl\'eaire et de Hautes Energies (LPNHE), Universit\'es Paris 6 et Paris 7, CNRS-IN2P3, France}
% ...
\affiliation[38]{Gran Sasso Science Institute (INFN), L'Aquila, Italy}
% ...
\affiliation[47]{INFN, Sezione di Torino, Italy}
\affiliation[48]{Osservatorio Astrofisico di Torino (INAF), Torino, Italy}
% ...
\affiliation[62]{Universidad Nacional Aut\'onoma de M\'exico, M\'exico}
\affiliation[63]{Institute for Mathematics, Astrophysics and Particle Physics (IMAPP), Radboud Universiteit, Nijmegen, Netherlands}
% ...
\affiliation[70]{Laborat\'orio de Instrumenta\c{c}\~ao e F\'\i{}sica Experimental de Part\'\i{}culas -- LIP and Instituto Superior T\'ecnico -- IST, Universidade de Lisboa -- UL, Portugal}
% ...
\affiliation[79]{Universidad de Santiago de Compostela, Spain}
% ...
\affiliation[c]{also at Vrije Universiteit Brussels, Brussels, Belgium}
% ...

\emailAdd{auger\_spokespersons@fnal.gov}

\abstract{
A search for ultra-high energy photons with energies above 1~EeV is performed using nine years of data collected by the Pierre Auger Observatory  in hybrid operation mode. 
An unprecedented separation power between photon and hadron primaries is achieved by combining measurements of the longitudinal air-shower 
     development with the particle content at ground measured by the fluorescence and surface detectors,  respectively. 
Only three photon candidates at energies 1~$-$~2 EeV are found, which is compatible with the expected
hadron-induced background.
Upper limits on the integral  flux of ultra-high energy photons of
0.038, 0.010, 0.009, 0.008 and 0.007 km$^{-2}$ sr$^{-1}$ yr$^{-1}$ are derived at 95\% C.L. for energy thresholds of 1, 2, 3, 5 and 10~EeV. These limits bound the fractions of photons in the all-particle integral flux below 0.14\%, 0.17\%, 0.42\%, 0.86\% and 2.9\%. 
For the first time the photon fraction at EeV energies is constrained at the sub-percent level. 
The improved limits are below the flux of diffuse photons predicted by some astrophysical scenarios for cosmogenic photon production.
The new results rule-out the early top-down models $-$ in which ultra-high energy cosmic rays are produced by, e.g., the decay of super-massive particles $-$ and challenge the most recent super-heavy dark matter models.
}

\begin{document}
\maketitle
\flushbottom

\section{Introduction}
\label{sect:intro}

Ultra-high energy (UHE) photons are among the possible particles contributing to the flux of cosmic rays. 
A flux of UHE photons is 
expected  from the decay of $\pi^0$  particles produced by protons interacting with the cosmic microwave background (CMB) in the so-called Greisen-Zatsepin-Kuz'min (GZK) effect~\cite{GZK1,GZK2}. The energy threshold of the process is  about 10$^{19.5}$~eV and photons are produced on average with around 10\% of the energy of the primary incident proton. 
The energy loss for the GZK protons limits their range to about a hundred Mpc: Only sources within this horizon contribute to the observed cosmic-ray flux above the GZK energy threshold producing a cut-off with respect to a continuation of the power-law energy spectrum. 
A flux suppression has been observed~\cite{HiResSpectrum,AugerSpectrum,AugerSpectrum2,TASpectrum} but the current experimental results are not sufficient to exclude other possible scenarios such as a limitation in the maximal acceleration energy of cosmic rays at the source. A combined fit of the energy spectrum and the mass composition measured by the Pierre Auger Observatory~\cite{NIM2015} - under simple assumptions on the astrophysical sources and on the propagation of cosmic rays - seems to favor the latter scenarios~\cite{DiMatteo}. Results obtained by the Telescope Array Collaboration prefer a GZK scenario when interpreting the observed mass composition as  proton-dominated up to the highest energies~\cite{TAFluxInterpretation}. This is however challenged with the limits on cosmogenic neutrino fluxes~\cite{Heinze:2015hhp,Aartsen:2016ngq,AugerNeutrino} and by the observed diffuse sub-TeV $\gamma$-radiation (see for example~\cite{Berezinsky2016}).
Within this context, the observation of GZK (or ``cosmogenic'') photons (and neutrinos) would be an independent proof of the GZK process. 
The expected flux of GZK photons is estimated to be of the order of 0.01-0.1\% depending on the astrophysical model (e.g., mass composition and spectral shape at the source)~\cite{Gelmini, Hooper, Kampert}.  

Moreover, a large flux of UHE photons is predicted in top-down models with ultra-high-energy cosmic rays (UHECR) originating from the decay of supermassive particles. Some of these models, severely constrained by previous experimental results on UHE photons~\cite{AugerPhotonSD,AugerPhotonHybrid,AugerPhotonHybridICRC,AugerPhotonSD2015}, have been recently re-proposed to accommodate the existing photon limits and to test the lifetime-and-mass parameter space of putative Super Heavy Dark Matter (SHDM) particles~\cite{AloisioSHDM}.
As opposed to neutrinos, photons undergo interactions with the extragalactic background light (EBL) inducing electromagnetic cascades, see e.g.~\cite{EleCa}. This makes photons sensitive to the extragalactic environment (e.g. EBL, magnetic fields). New physics scenarios (e.g., violation of Lorentz invariance, photon-axion conversion) related to interaction or propagation effects can also be tested with photons and neutrinos (see for example~\cite{LIVphotons,LIVphotons2,LIVneutrinos,axions}).  

The production of UHE photons  at astrophysical sources accelerating high-energy hadrons has been tested performing a blind search for excesses of photon-like events over the sky exposed to the Pierre Auger Observatory~\cite{AugerDirectionalPhotons} and  searching for a correlation with the directions of targeted sources~\cite{AugerTargetPhotons}. These analyses consider events in the energy region between 10$^{17.3}$ and 10$^{18.5}$ eV. The reported null results set bounds to the photon flux emitted by discrete sources and on the extrapolation of $E^{-2}$ energy spectra of TeV  (1 TeV = 10$^{12}$ eV) gamma-ray sources within or near the Galaxy. 

No photons with energies  above 1 EeV (10$^{18}$~eV) have been definitively identified so far, bounding  their presence in the cosmic-ray flux to less than a few percent. 
Two analyses have been conducted by the Pierre Auger Collaboration in previous work, each one optimizing the energy range to the sensitivity of the two independent detectors comprising the Observatory.
The photon detection efficiency of the surface detector enables a photon search with large event statistics at energies above 10~EeV~\cite{AugerPhotonSD}. The analysis, recently updated in~\cite{AugerPhotonSD2015}, constrains the integral photon flux to less than $1.9\times 10^{-3}$, $1.0\times10^{-3}$ and $4.9\times10^{-4}$~km$^{-2}$~sr$^{-1}$~yr$^{-1}$ above 10, 20 and 40~EeV, respectively. 
A second analysis based on the detection of air-showers with the fluorescence telescopes operating in hybrid mode extended the energy range down to 2~EeV with statistics lower by a factor 10 because of the detector duty cycle~\cite{AugerPhotonHybrid}. In that work the identification of photon-induced air showers relied on the measurement of the depth of the air-shower maximum for a  sub-sample of hybrid events geometrically constrained to ensure a composition-independent detection efficiency.  Upper limits were placed on the integral photon fraction of 3.8\%, 2.4\%, 3.5\% and 11.7\%  above 2, 3, 5  and 10~EeV, respectively. 
A novel approach, combining the shower maximum observed by fluorescence telescopes and the signal at ground measured by the surface detectors is presented here. With respect to~\cite{AugerPhotonHybrid}, the data set is updated adding six more years of data and the improved background rejection and the use of a less stringent data selection allow one to achieve for the first time the sensitivity required to explore photon fractions in the all-particle flux down to 0.1\% and to extend the search for photons to 1~EeV. 

The paper is organized as follows. After a brief description of the Pierre Auger Observatory (section~\ref{sect:Auger}), the observables sensitive to the electromagnetic and hadronic nature of extensive air showers (EAS) are introduced in section~\ref{sect:observables}. The analysis is applied to 9 years of high-quality selected data as discussed in section~\ref{sect:dataset}. The multi-variate analysis tuned to identify photon-like events is described in section~\ref{sect:analysis}. In the absence of any significant signal, upper limits on the integral photon flux are derived. Results and systematic uncertainties are reported in section~\ref{sect:results}. 
A discussion is given in section~\ref{sect:conclusions} of constraints on astrophysical
and exotic models for the origin of UHECRs along with expectations of
more sensitive searches for UHE photons in the future.

\section{The Pierre Auger Observatory} \label{sect:Auger}
The Pierre Auger Observatory is located in Malarg\"ue, Argentina, and consists of a surface detector (SD) array of 1660 water Cherenkov  stations deployed over a triangular grid of 1.5~km spacing and covering an area of 3000~km$^2$. The stations sample the density of the secondary particles of the air shower at the ground and are  sensitive to  the electromagnetic, muonic and hadronic components.  
The Cherenkov light produced in the water volume of the station is collected by three photo-multiplier tubes (PMTs) and measured in units of VEM (Vertical Equivalent Muon, i.e. the signal produced by a muon traversing the station vertically). The signals are acquired and sent to the central acquisition system if they are above a threshold of 1.75 VEM in the three PMTs or if they match the time-over-threshold (ToT) algorithm requirements of at least 13 time bins  above a threshold of 0.2 VEM in a 3 $\mu$s window for at least two PMTs. 
The threshold trigger selects large signals, not necessarily spread in time, and is mostly effective for the detection of inclined showers for which only the muonic component reaches the ground. On the other hand, the ToT trigger selects signals spread in time and is thus more efficient for events with arrival directions closer to the zenith ~\cite{SDtrigg}.

The SD array is overlooked by 27 telescopes grouped in 5 buildings forming the fluorescence detector (FD)~\cite{FDpaper}.  The FD observes the longitudinal development of the shower by detecting the fluorescence and Cherenkov light emitted during the passage of the secondary particles of the shower in the atmosphere.  
Unlike the SD, the fluorescence telescopes work only during clear and moonless nights, for an average duty cycle of about 14\%~\cite{exposure2010}.

The presence of aerosols and clouds alters the intensity of light collected by the telescopes, the FD trigger efficiency and the observed longitudinal profile. 
Several monitoring systems are installed to measure the aerosol content and the cloud coverage. The vertical aerosol optical depth (VAOD) is measured using two lasers deployed at the center of the array (the Central Laser Facility, CLF, and the eXtreme Laser Facility, XLF)~\cite{aerosol,aerosol2,CLF1,CLF2}. Close to each FD site, a lidar system~\cite{Lidar} provides a cross-check of the aerosol content and measures the coverage and height of the clouds. In addition, the cloud coverage for each pixel of the FD is inferred from the analysis of the images acquired by the infrared cameras installed on the roof of the FD buildings~\cite{clouds}.  

If at least one SD station detects a signal in time and spatial coincidence with the FD, a hybrid reconstruction can be performed~\cite{FDpaper}. 
In the hybrid mode the geometry of the event is determined from the arrival time of the light at the FD pixels with the additional constraint provided by the timing information from the SD.
The longitudinal profile  is then reconstructed taking into account the scattering and absorption of light from the shower axis to the telescope. 
It is the main measurement for determining the energy of the primary cosmic ray and constraining its mass~\cite{Xmax}.
The depth, \Xm,  at which the shower reaches its maximum development is directly derived from the fit of a Gaisser-Hillas function~\cite{Gaisser-Hillas} to the longitudinal profile of the
air shower. The parameter \Xm is well known to be anti-correlated with the mass of the primary cosmic ray at any fixed energy. 
The total energy of the primary particle is determined from the integral of the fitted Gaisser-Hillas function corrected for the invisible energy~\cite{Mariazzi} carried by penetrating particles (mostly neutrinos and muons). The correction is about 1\% for electromagnetic showers and 10-15\% for nuclear primaries  depending only weakly on the primary mass and on choice of the hadronic interaction models. 

Unless differently specified, in this paper the photon energy $E_\gamma$ is used as default for simulations and data, independently of the nature of the primary particle.
\begin{figure}[t]
\centering
\includegraphics[width=0.49\textwidth]{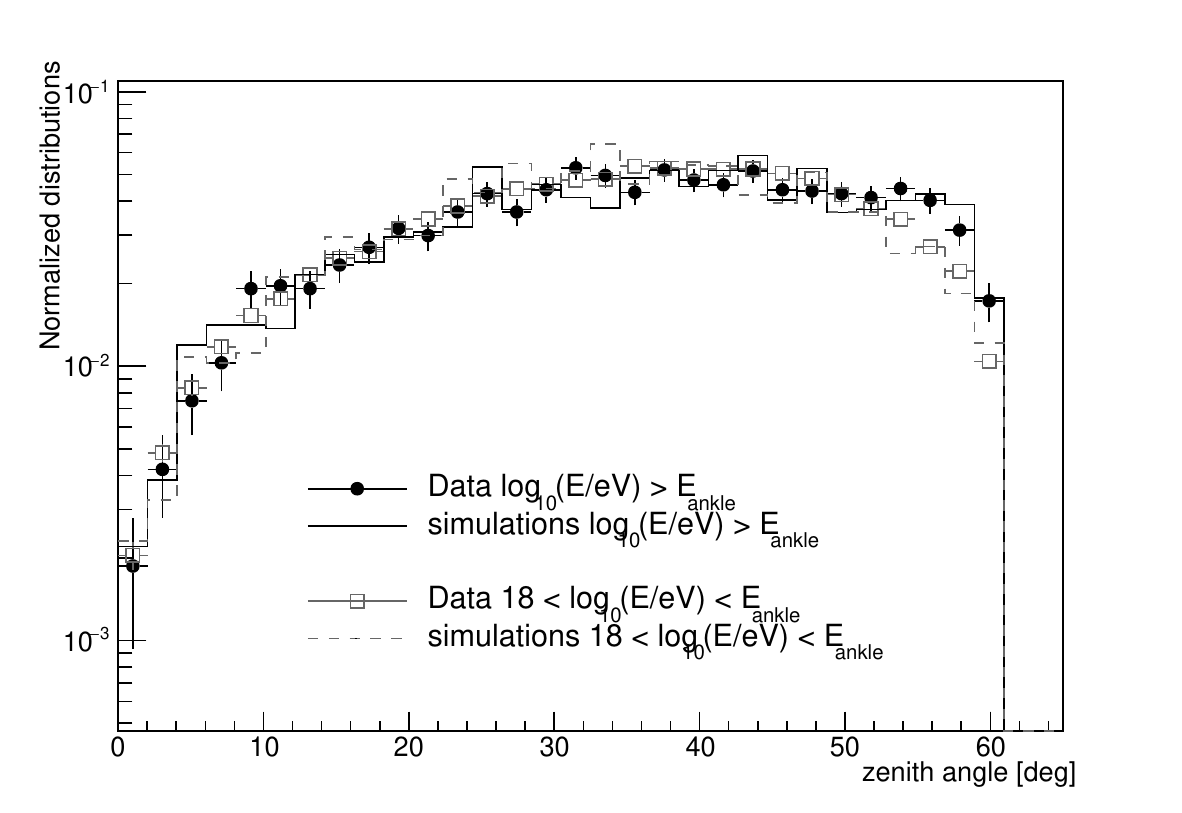}
\includegraphics[width=0.49\textwidth]{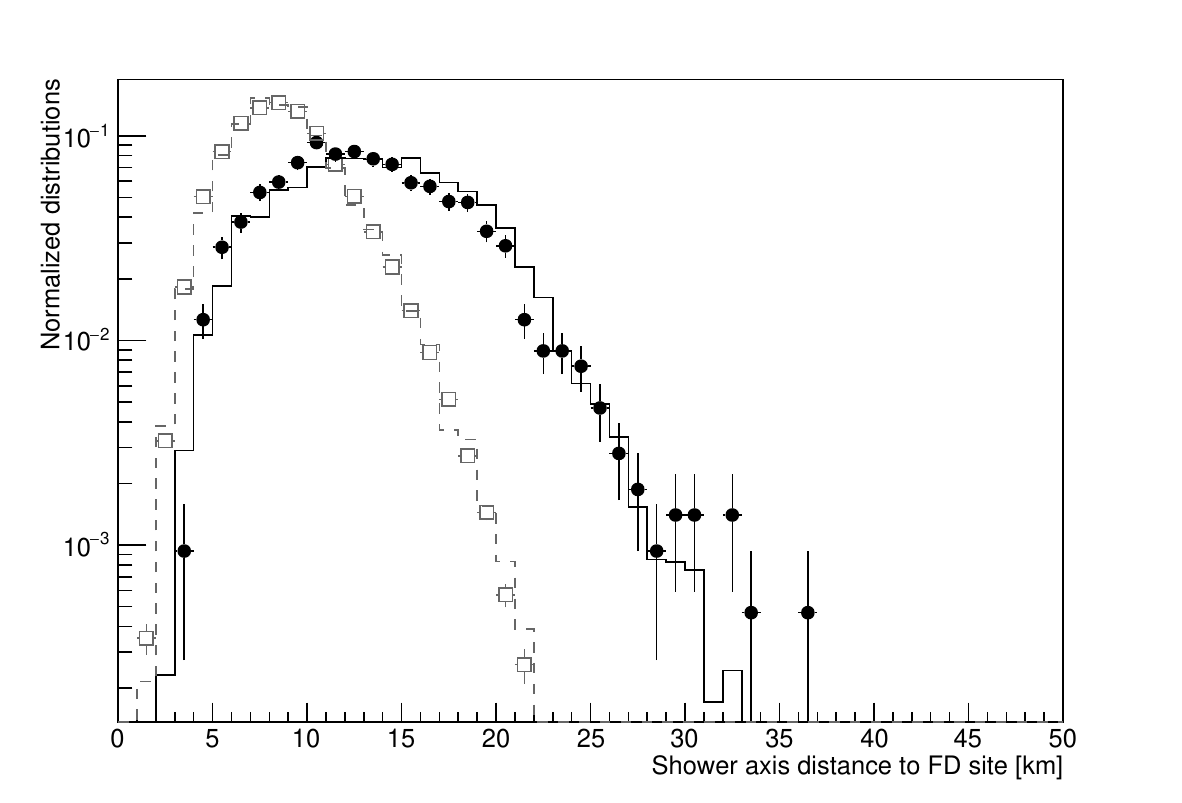}
\caption{ Distributions of the zenith angle (left) and the distance of the shower axis to the FD (right) are shown as examples of the agreement between time-dependent simulation (histograms) and data (markers) in two separate energy intervals (below and above the ``ankle'' spectral feature $E_{\rm{ankle}} \simeq 10^{18.68}$~eV~\cite{AugerSpectrum2}). Events in data and simulations are selected applying the criteria described in section~\ref{sect:dataset}, with the exception of the energy cut. Simulations are re-weighted according to the spectral index given in~\cite{AugerSpectrum2} and a mixed composition (50\% proton - 50\% iron) is assumed.}\label{fig:dataMC}
\end{figure}

\section{Observables for the photon search}\label{sect:observables}

The search for UHE photon primaries is based on the different development and particle content of electromagnetic and hadronic air-showers. The induced electromagnetic cascades develop slower than hadronic ones so that \Xm is reached closer to the ground. Proton and photon simulated showers have average
Xmax values that differ by about 200~g/cm$^{2}$ in the EeV energy range. This difference is enhanced at energies above 10$^{19}$~eV because of the Landau-Pomeranchuk-Migdal (LPM) effect~\cite{LPM1,LPM2}. 
At higher energies, above 50~EeV, photons have a non-negligible probability to convert in the geomagnetic field~\cite{preshower1,preshower2,Homola2007} producing a bunch of low-energy electromagnetic particles, called ``pre-shower'', entering the atmosphere. The \Xm of the pre-showered cascades is smaller than for non-converted ones and the separation between the average \Xm for photons and proton primaries is reduced. 

The shower development and the nature of the primary cosmic ray determine the content and the shape of the distribution of particles at ground as a function of the distance from the shower axis (Lateral Distribution Function, LDF).  Photon-induced showers generally have a steeper LDF compared to hadron primaries because of the sub-dominant role played by the flatter muonic component. The high-energy effects (LPM and pre-showering) do not affect the muon content, however the different stage of shower development (i.e., \Xm) leads to a modification of the observed LDF.  Given the steeper LDF and the muon-driven SD triggers, the footprint at the ground, and consequently the number $N_{\rm{stat}}$ of triggered stations, is typically smaller for electromagnetic showers~\cite{LTP}. 
These features are combined in the observable $S_b$~\cite{Ros}: 
\begin{equation} 
S_b = \sum_{i}^N S_i \left( \frac{R_i}{R_0} \right)^{b}
\end{equation}
where $S_i$ and $R_i$ are the signal and the distance from the shower axis of the $i$-th station, $R_0 = 1000$~m is a reference distance  and $b = 4$ is a constant  optimized to have the best separation power between photon and nuclear primaries in the energy region above 10$^{18}$~eV. 

\begin{figure}[t]
\centering
\includegraphics[width=0.96\textwidth]{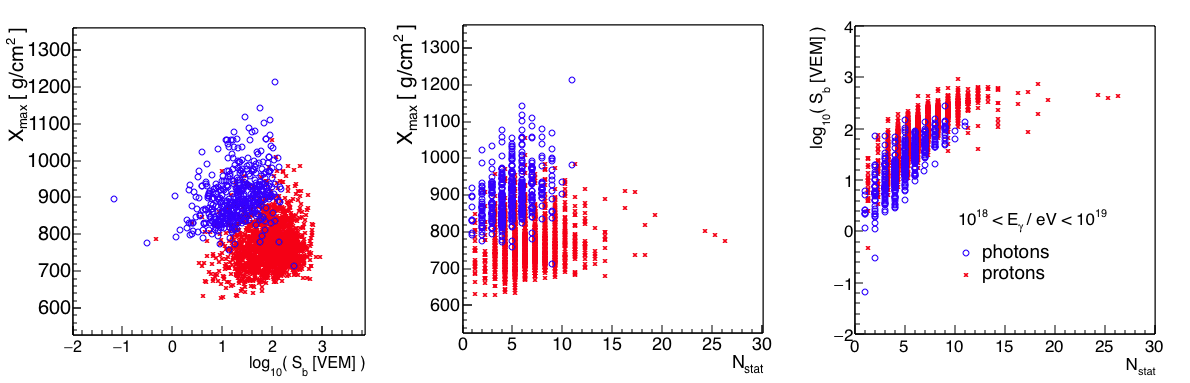}
\caption{Correlation between the discriminating observables used in the multivariate analysis for the energy range $10^{18} < E_{\gamma} < 10^{19}$~eV: the red stars and the blue circles are the proton and photon simulated events, respectively.  Events are selected applying the criteria in section~\ref{sect:dataset}. For a better visibility of the plot only 5\% of events are plotted and a shift of 0.25 is applied to $N_{\rm{stat}}$ for proton events.}\label{fig:scatter}
\end{figure}

Detailed simulations of the air-showers and of the detector response have been performed to study the photon/hadron discrimination. A data set of about 60000 photon-induced showers have been generated with CORSIKA version 6.990~\cite{CORSIKA} with energy between 10$^{17}$~eV and 10$^{20}$~eV following a spectrum $E^{-1}$ in bins of 0.5 in the logarithm of energy.  Events are sampled from an isotropic distribution, with the zenith angle $\theta$ ranging between 0 and 65 degrees. The azimuth angle $\phi$ is uniformly distributed between 0 and 360$^\circ$. Pre-showering and LPM effects are included in the simulations.  Proton and iron showers are simulated with CORSIKA version 7.4002 adopting the most up-to date hadronic interaction models, EPOS LHC~\cite{epos-lhc} and QGSJET-II-04~\cite{qgsjet2-04}. A total of 25000 showers have been generated for each hadronic model and primary type.  Each shower is  resampled 5 times, each time with a different impact point at ground uniformly distributed within an area enclosing the array and a border such that the trigger efficiency of each surface station is less than 1\% outside it~\cite{LTP}.  Events are processed through the Offline software~\cite{Offline} which includes a detailed simulation of the FD and the light propagation from the shower to the FD camera and a Geant4-based~\cite{Geant4} simulation of the SD. A time-dependent approach developed for the energy spectrum in~\cite{AugerSpectrum2012} is used for a realistic estimate of the detection efficiency and the discrimination performance. 
In this approach, the actual status of the FD and the SD, as well as the atmospheric conditions, are taken into account and the events are distributed according to the on-time of the hybrid detector. 
As validations of the procedure, Fig.~\ref{fig:dataMC} demonstrates the comparison between data and simulations for two reconstructed observables (zenith angle, left,  and the shower-axis distance from the telescope, right) in two energy intervals. 
Fig.~\ref{fig:scatter} shows the correlation between the discriminating observables \Xm, $S_b$, $N_{\rm{stat}}$  for selected samples of well  reconstructed photons (blue circles) and protons (red stars) events, the latter ones being the main source of background for this study. 

\section{Data set}\label{sect:dataset}
\begin{table}[!t]
 \begin{center}
  \begin{tabular}{lrr}
   \hline
   \hline
    Criteria & N events  & efficiency [\%] \\
   \hline
Trigger  & 3306730 &  -- \\
Detector & 1490335 & 45.07\\
Geometry & 610192 & 40.94\\
Profile & 62776 & 10.29\\
$E_{\gamma} > 10^{18}$~eV  & 18968 & 30.22\\
$S_{b}$ & 17297 & 91.19\\
Atmosphere & 8178 & 47.28\\
   \hline
   \hline
  \end{tabular}
 \end{center}
 \caption{Event selection criteria, number of events after each cut and selection efficiency with respect to the previous cut.}
 \label{tab:eff}
\end{table} 

The analysis presented in this work uses hybrid data collected between January 2005 and December 2013. Selection criteria are applied to ensure a good geometry and profile reconstruction and a reliable measurement of the discriminating observables. These cuts are detailed below. 

\paragraph*{{\bf Trigger and detector levels.}}
The initial data set (trigger level) consists of all events passing the very loose trigger requirements of the data acquisition~\cite{FDpaper}. Consequently it includes a fraction of events that are not due to air-shower events (e.g. lightning or low energy events with a random-coincidence station) and are thus discarded. 
Data periods without good FD or SD working conditions, mostly during the  construction phase of the observatory (e.g., camera calibrations in the FD and unstable conditions of the SD trigger) are rejected. 

\paragraph*{{\bf Geometry level.}} 
The station selected in the hybrid reconstruction is required to be within 1500~m of the shower axis and its timing has to be within 200~ns of the expected arrival time of the shower front~\cite{FDpaper}.  Hybrid events with  a successful reconstruction of the shower axis ($\chi^2$ of the temporal fit has to be smaller than 7) and with a zenith angle up to 60$^\circ$ are considered. More inclined events are not included in this analysis because of the absorption of the electromagnetic components of the EAS in the atmosphere and the resultant small trigger efficiency for photons at low energies. 
 As a quality selection criterion, the angular track length, defined as the angular separation between the highest and lowest FD pixels in the track, is required to be larger than 15$^{\circ}$. A resolution better than 50~m on the core position and of 0.6$^\circ$ on the arrival direction are obtained with these cuts for events with energy above 10$^{18}$~eV.  
Events are selected if they land within a fiducial distance from the telescope for which the FD trigger efficiency is flat within 5\% when shifting the energy scale by $\pm$14\%~\cite{AugerSpectrum2012}. 
This distance, parameterized in different energy intervals, is based on simulations and is mostly independent of the mass composition and hadronic models. It is around 14~km at 10$^{18}$~eV and 30~km at 10$^{19}$~eV.

\paragraph*{{\bf Profile level.}} 
For a reliable measurement of the \Xm and of the energy, the goodness of the Gaisser-Hillas fit is tested requiring a reduced $\chi^2$ smaller than 2.5. The request of a viewing angle between the shower axis and the telescope larger than 20$^\circ$ rejects events pointing toward the FD and having a large Cherenkov light contamination. To avoid biases in the reconstruction of the longitudinal profile, the \Xm has to be observed in the field of view of the telescope and gaps in the profile have to be shorter than 20\% of the total observed length. To reject events with a flat  profile, for which the \Xm determination is  less reliable, the ratio between the $\chi^{2}$ of a Gaisser-Hillas and a linear fit of the profile is required to be smaller  than 0.9~\cite{AugerPhotonHybrid}.  Events are selected if the relative uncertainty on the reconstructed energy is smaller than 20\%. 
These criteria ensure an energy resolution between 10 and 15\% improving with energy and an  
\Xm resolution from about 20~g/cm$^{2}$ at 10$^{18}$ eV to about 15~g/cm$^2$ above 10$^{19}$~eV. 

\paragraph*{{\bf $S_b$ selections.}} 
 Artificially small values of $S_b$ and $N_{\rm{stat}}$ can be obtained for events landing in region of the array close to the borders or with incomplete station deployment (during the construction phase of the Observatory) or having stations inactive because of temporary detector inefficiencies. To reject these events,  which would mimic photon candidates, at least 4 active stations are required within the first 1500~m hexagon around the station with the largest signal. This criterion rejects 9\% of the events. 

\paragraph*{{\bf Atmosphere.}} To minimize biases from possible distortions of the longitudinal profile produced by clouds, a measurement of the cloud coverage by infrared camera or by the lidar system is required to be available and to be lower than 25\%.  Time periods without information on the aerosol content of the atmosphere or with poor viewing conditions are excluded requiring that the measured vertical aerosol optical depth (VAOD), integrated from the ground to 3 km, is smaller than 0.1. \\

The selection efficiencies with respect to the full set of recorded events are given in  table~\ref{tab:eff}.
The final data set  among which photon candidates are searched for contains 8178 events with energy $E_\gamma$ larger than 10$^{18}$~eV.   

\begin{figure}[!t]
\centering
\includegraphics[width=0.48\textwidth]{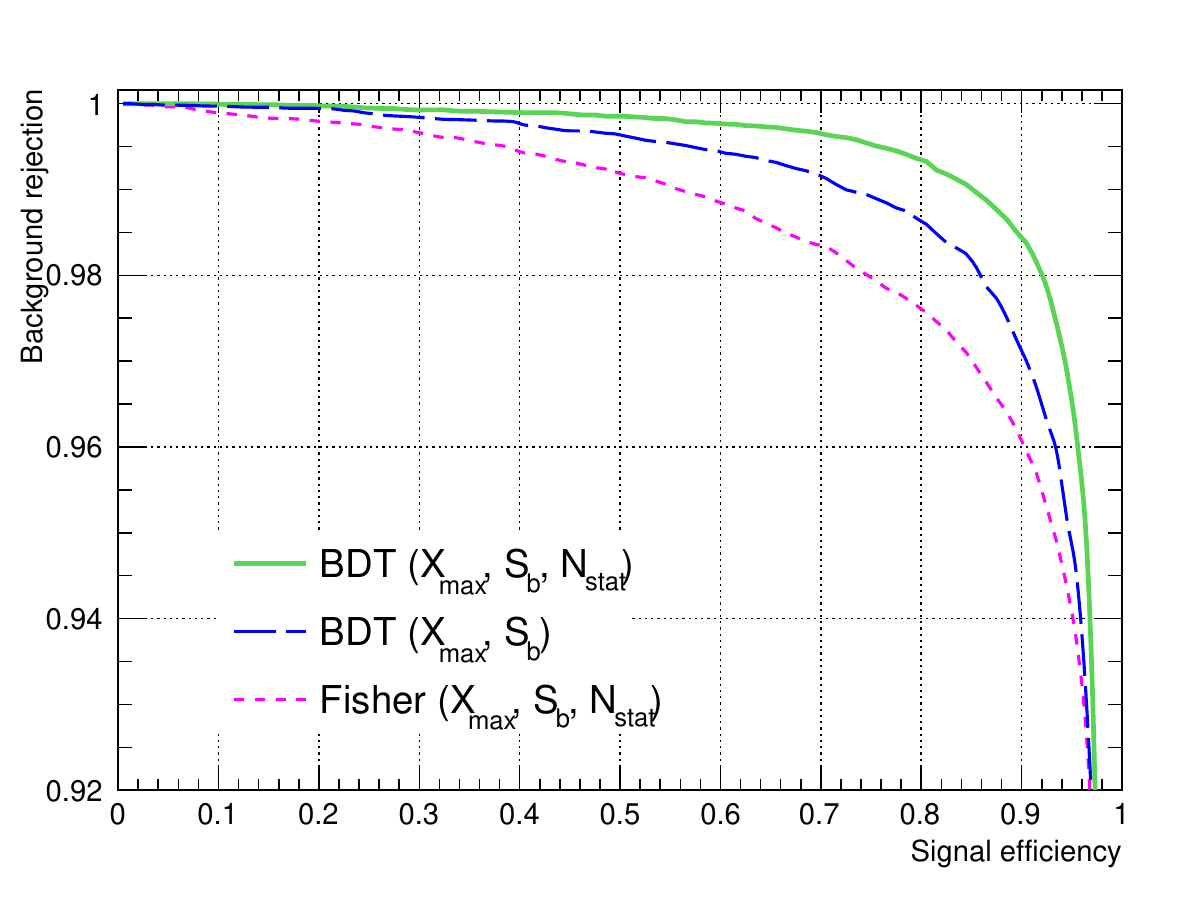}
\includegraphics[width=0.505\textwidth]{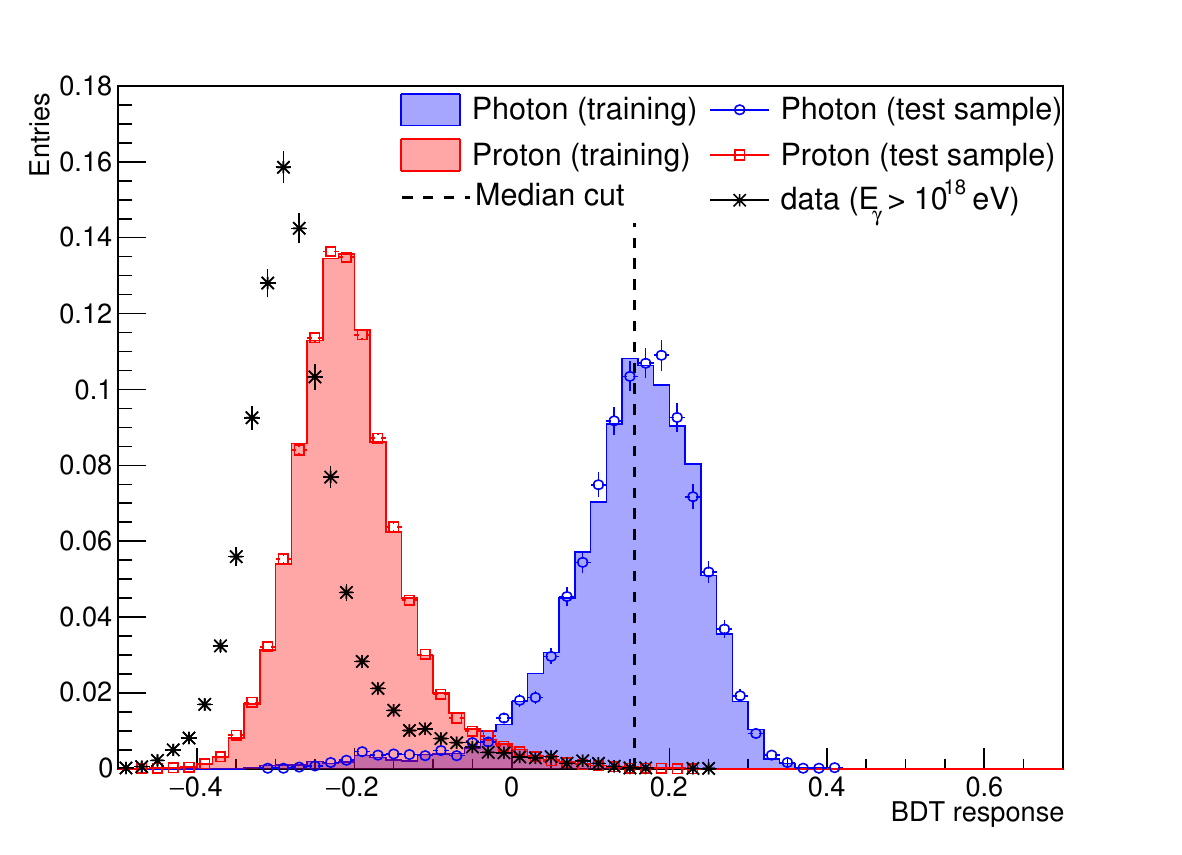}
\caption{Left: curve of the background rejection efficiency against the signal efficiency for different algorithms and observables.  Right: distribution of the Boosted Decision Tree observables for signal (photon, blue), background (proton, red) and data (black). For simulations both the training and the test samples are shown. The cut at the median of the photon distribution is indicated by the dashed line. QGSJET-II-04 used as high-energy hadronic interaction model.}\label{fig:MVA}
\end{figure}

\section{Analysis}\label{sect:analysis}
To identify a possible photon signal among the large background due to hadronic primaries, a multivariate analysis is performed adopting different algorithms.  The Boosted Decision Tree (BDT) has been found to provide the best separation. This method has also the advantage of being more stable against the inclusion of observables with weak discriminating power.  The variable ranking gives \Xm as the strongest variable followed by $S_b$ and $N_{\rm{stat}}$. To take into account the energy and angular dependences of these three observables, the energy and zenith angle are included in the multivariate analysis. 
A test excluding the least significant discriminating observable, $N_{\rm{stat}}$, has been performed to evaluate its impact on the separation power. The  background rejection versus  signal efficiency for the BDT using all observables and for the case excluding $N_{\rm{stat}}$ are drawn in Fig.~\ref{fig:MVA} (left). For a photon selection efficiency $\epsilon_\gamma$~=~50\% the use of $N_{\rm{stat}}$ reduces the background contamination by more than a factor 2, from 0.37\% to 0.14\%. Thus the analysis is performed considering all discussed observables. 
In the preliminary analysis presented in~\cite{AugerPhotonHybridICRC}, a Fisher method trained only with \Xm and $S_b$ and optimized in three different energy ranges was adopted for the  sake of simplicity. For comparison, the performance of the Fisher algorithm is also illustrated in Fig.~\ref{fig:MVA} (left). The background rejection efficiency is found to be around 99\% for $\epsilon_\gamma=$~50\%. 
In the multivariate analysis events are weighted according to a power law spectrum $E^{-\Gamma}$ with $\Gamma = 2$. The performance of the BDT (using all the discriminating observables) has been tested against the  variation of the spectral index. For a simulated flux with $\Gamma = 1.5$ and $\Gamma = 2.5$, the background contamination at 50\% of the photon efficiency is 0.07\% and 0.24\%, respectively (cfr. 0.14\% obtained in the case $\Gamma = 2$). These results are expected due to the larger (smaller) contribution of the highest energy events for which \Xm and $S_{b}$ have better separation. 

The BDT response is given in Fig.~\ref{fig:MVA} (right) for data and for photon and proton QGSJET-II-04 simulations. 
The discrepancy between the data and the proton simulations  is in agreement with the current experimental indications of a composition varying from light to heavier composition in the EeV range~\cite{Xmax, XmaxDistrib, XmaxS1000} and the muon deficit observed in simulations with respect to the Auger data~\cite{Farrar,muon_horiz}.   
To identify photons, a cut is defined at the median of the BDT response distribution for photons. This way, the signal efficiency remains constant independently of the composition and hadronic model assumptions. Events having a BDT response larger than the median cut (dashed vertical line in Fig.~\ref{fig:MVA}, right) are selected as ``photon candidates''.  
A  background contamination of $\sim$~0.14\% is obtained for proton showers using QGSJET-II-04 and it becomes $\sim$~0.21\% when the EPOS LHC model is used. This background level overestimates the one  expected in data because of the composition and muon arguments discussed above. 
As a reference, the multivariate analysis has been performed providing a mixture of 50\% proton and 50\% iron as input to the training phase. The background contamination in this case reduces to $\sim$~0.04\% with the main contribution coming from the smaller values of \Xm. For the available data set, this background contamination corresponds to 11.4 (3.3) events in the case of a proton (mixed) composition, assuming the QGSJET-II-04 model. 

\begin{figure}[t]
\centering
\includegraphics[width=0.49\textwidth]{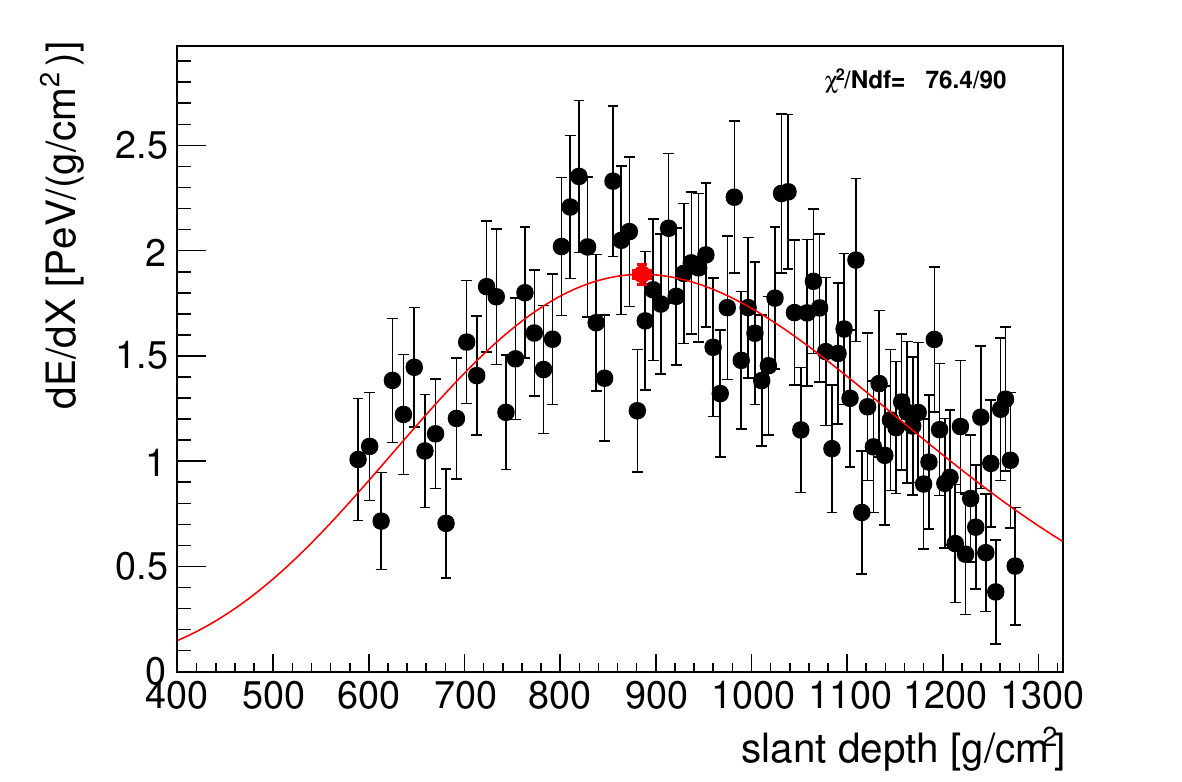}
\includegraphics[width=0.49\textwidth]{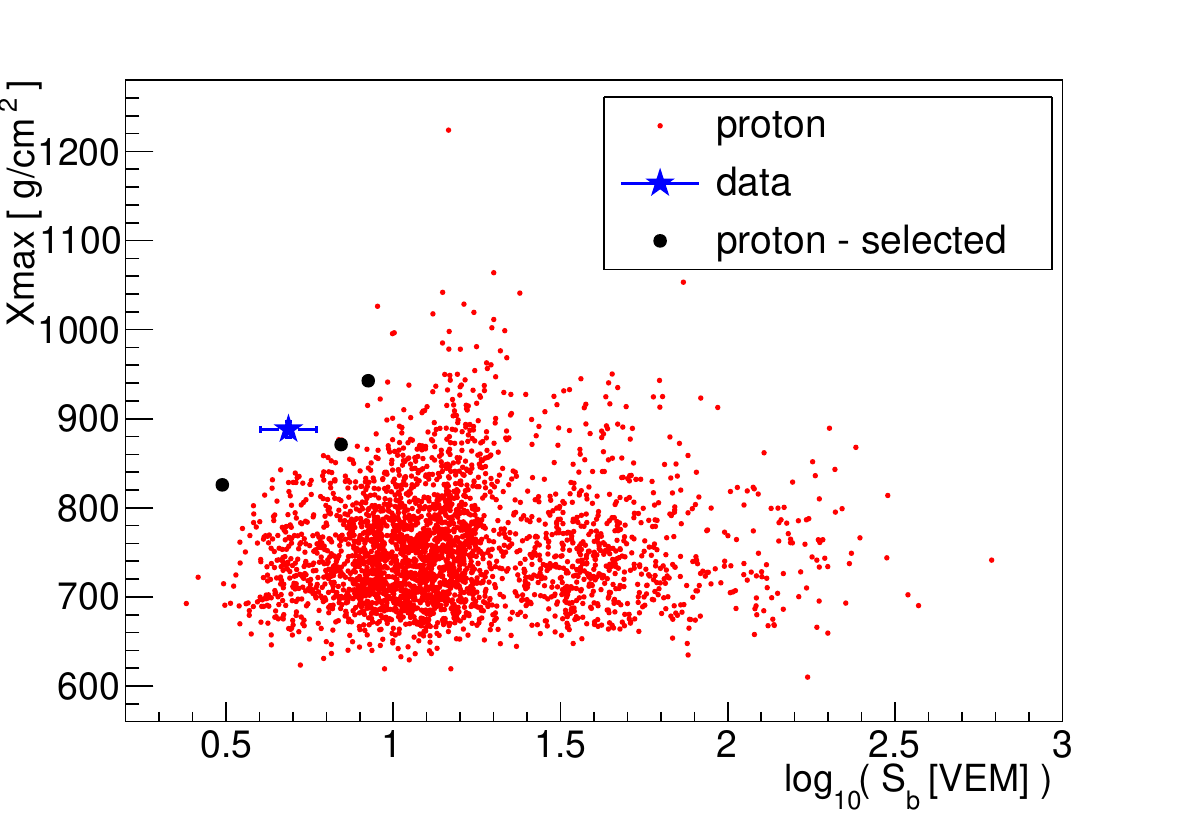}
\caption{Left: Longitudinal profile and Gaisser-Hillas fit of one of the selected photon candidates (ID 6691838). Right: Correlation plot of \Xm and $S_b$ for the candidate (blue star) and  dedicated proton events simulated with the same energy, geometry and detector configuration as the real event (red dots). Three out of 3000 simulated proton showers are selected as photon candidates (black circles).  }\label{fig:cand}
\end{figure}
\begin{table}
\begin{center}
{\small
  \begin{tabular}{lccccccc}
   \hline
   \hline
    Event ID  & $E_{\gamma}$ [EeV]  & Zenith [$^\circ$] &  \Xm [g/cm$^{2}$] & $S_b$ [VEM] & $N_{\rm{stat}}$ & $l$ [$^\circ$] & $b$ [$^\circ$]\\
   \hline
    3218344   &   1.40$\,\pm\,$0.18  & 34.9$\,\pm\,$0.9   & 851$\,\pm\,$31  & 2.04$\,\pm\,$0.77 & 2  & 218.21$\,\pm\,$1.29 & -25.67$\,\pm\,$0.36 \\ 
    6691838   &   1.26$\,\pm\,$0.05  & 53.9$\,\pm\,$0.3   & 886$\,\pm\,$9    & 4.94$\,\pm\,$1.21 & 2 & 100.45$\,\pm\,$0.57  &  -46.25$\,\pm\,$0.25 \\ 
    12459240 &  1.60$\,\pm\,$0.14    & 49.4$\,\pm\,$0.4   & 840$\,\pm\,$21  & 9.57$\,\pm\,$2.56 & 3 & 324.94$\,\pm\,$0.37 & -24.70$\,\pm\,$0.60 \\ 
    \hline
   \hline
  \end{tabular}
  }
  \end{center}
  \caption{List of the events selected as photon candidates with the main quantities used for photon-induced air-showers identification and with their arrival directions in galactic coordinates ($l$,$b$). }
    \label{tab:candidates}
  \end{table}
The BDT analysis is applied to the full data set described in section~\ref{sect:dataset}. After the selection 8178, 3484, 2015, 983 and 335 events are left for the analysis above 1, 2, 3, 5 and 10 EeV, respectively. 
Three events pass the photon selection cuts and all of them are in the first energy interval ($1-2$ EeV), close to the energy threshold of the analysis. This number of events is compatible with the expected nuclear background. 
Details of the candidate events are listed in table~\ref{tab:candidates}. The arrival directions of the three photon-like events have been checked against a catalogue of astrophysical sources of UHECRs whose distance is limited to a few Mpc because of UHE photons interaction on the extragalactic background radiation~\cite{AugerTargetPhotons}. The smallest angular distances between the candidates and any of the objects in the catalogue is found to be around 10$^{\circ}$. 
One candidate (ID 6691838) was also selected in a previous analysis~\cite{AugerPhotonHybridICRC}. Its longitudinal profile is shown in Fig.~\ref{fig:cand} (left). In Fig.~\ref{fig:cand} (right), the values of  \Xm and $S_b$ for this event are compared to the measured ones in dedicated simulations having the same geometry and energy of this event. In the data sample of simulated protons, three out of 3000 showers pass the photon selections and are misclassified, in agreement with the expected average background contamination.

\begin{figure}
\CenterFloatBoxes
  \begin{floatrow}
  \hfill
    \ffigbox{%
       \vspace{0.5cm}
       \centering
        \includegraphics[width=0.5\textwidth]{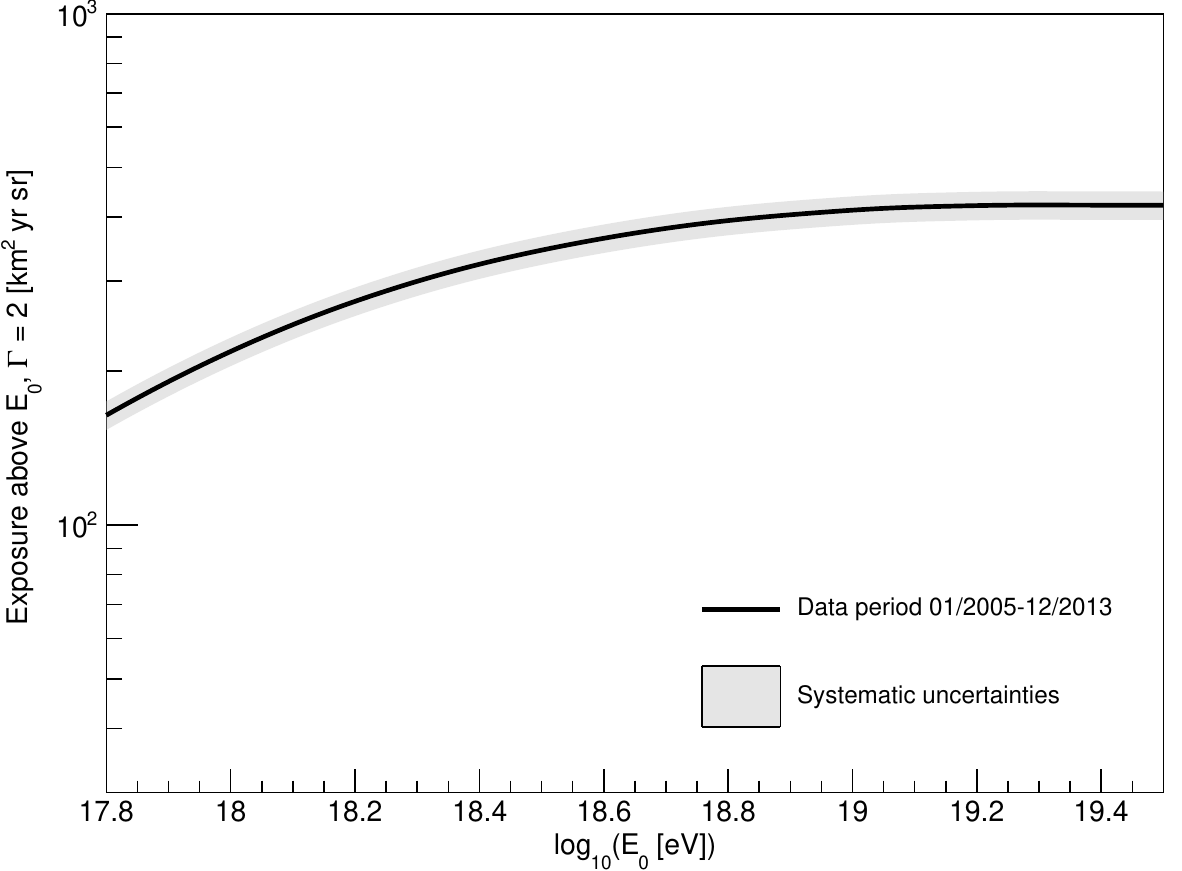}
    }{%
    \centering
      \captionof{figure}{Hybrid exposure for primary photons in the
        time interval 1 January 2005 - 31 December 2013, assuming a
        power-law spectrum with $\Gamma = 2$. Systematic uncertainties due to the ontime and the trigger efficiency are shown as a gray band.}%
      \label{fig:expo}%
    }
     \hspace{0.1cm}
     \vspace{0.4cm}
     \capbtabbox{%
       \begin{tabular}[width=0.4\textwidth]{llr}
   \toprule
   \multicolumn{3}{c}{\small{Detector systematic uncertainties} }\\
   \midrule
    \small{Source}            &   \small{Syst. uncert.}  & UL$^{0.95}$ change \\ 
   &  &   (E$_\gamma >$ 1 EeV)\\
    \midrule 
    \small{Energy scale}   &   $\pm$ 14\%                 & (+18, -38)\%  \\ 
    \small{\Xm scale}       &   $\pm$ 10 g/cm$^{2}$  & (+18, -38)\% \\ 
    \small{S$_{b}$}           &  $\pm$ 5\%                    & (-19, +18)\% \\
    \small{Exposure}         &  $\pm$ 6.4\%                & (-6.4, +6.4)\%\\
        \bottomrule
  \end{tabular}
    }
 {%
    \vspace{0.5cm}
      \caption{Relative changes of the upper limits on the photon flux for different sources of systematic uncertainties related to the detector. Only the first energy bin ($E_\gamma > 1$~EeV) is  reported as the mostly affected one. }%
      \label{tab:syst}%
      }
  \end{floatrow}

 \end{figure}

\section{Results}\label{sect:results}
Since the number of selected photon candidates is compatible with the background expectation, upper limits (UL) on the integral photon flux at 95\% confidence level (C.L.) are derived as: 

\begin{equation}\label{eq:UL}
\Phi_{UL}^{0.95} (E_{\gamma}>E_0)= \frac{N_{\gamma}^{0.95} (E_{\gamma} > E_0)}{\mathcal{E_{\gamma}}(E_{\gamma}>E_0 | E_{\gamma}^{-\Gamma})} 
\end{equation}

where $N^{0.95}_{\gamma}$ 
is the Feldman-Cousins upper limit at 95\% CL on the number of photon candidates assuming zero background events and $\mathcal{E_{\gamma}}$ is the integrated exposure above the energy threshold $E_0$, under the assumption of a power law spectrum $E^{-\Gamma}$ (if not differently stated $\Gamma = 2$ as in previous publications~\cite{AugerPhotonSD}): 
\begin{equation}\label{eq:exposure}
\mathcal{E_{\gamma}}  = \frac{1}{c_E}\int_{E_{\gamma}}\int_{T}\int_{S} \int_{\Omega}E_{\gamma}^{-\Gamma} \epsilon(E_{\gamma},t,\theta,\phi,x,y)\,dS\,dt\,dEd\Omega
\end{equation}
with $\epsilon$ being the overall efficiency for photons as a function of energy ($E_{\gamma}$),  time ($t$), zenith angle ($\theta$), azimuth ($\phi$) and position (x,y) of the impact point at ground. $c_E$ is a normalization coefficient: $c_E = \int E^{-\Gamma} dE$. $\Omega$ is the solid angle and the area $S$ encloses the array and corresponds to the generation area used for the simulations. 
The hybrid exposure after photon selection criteria is shown in Fig.~\ref{fig:expo} (left). 

\begin{figure}[t]
\centering
\includegraphics[width=0.7\textwidth,]{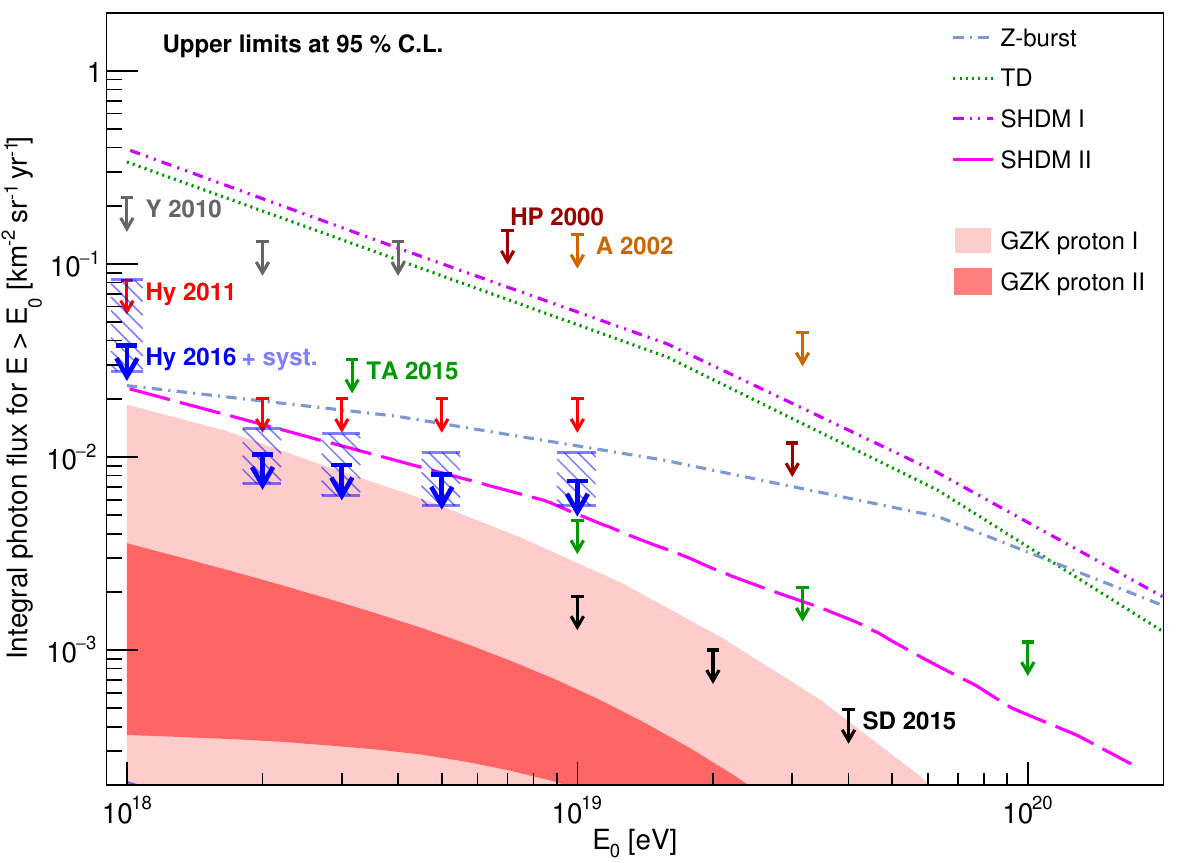}
\caption{Upper limits on the integral photon flux derived from 9 years of hybrid data (blue arrows, Hy~2016) for a photon flux E$^{-2}$ and no background subtraction. The limits obtained when the detector systematic uncertainties are taken into account are shown as horizontal segments (light blue) delimiting a dashed-filled box at each energy threshold.  Previous limits from Auger: (SD~\cite{AugerPhotonSD2015} and Hybrid 2011~\cite{AugerPhotonHybridICRC}), for Telescope Array (TA)~\cite{TAphotons}, AGASA (A)~\cite{Agasa}, Yakutsk (Y)~\cite{Yakutsk} and Haverah Park (HP)~\cite{HaverahPark} are shown for comparison. None of them includes systematic uncertainties. The shaded regions and the lines give the predictions for the GZK photon flux~\cite{Gelmini,Kampert} and for top-down models (TD, Z-Burst, SHDM I\cite{SHDM2} and SHDM II~\cite{AloisioSHDM}). }\label{fig:limits} 
\end{figure}

Using equation~\ref{eq:UL} and the analysis trained on photon and
proton QGSJET-II-04 simulations, with spectral index $\Gamma = 2$,
upper limits to the integral photon flux are set to 0.038, 0.010, 0.009, 0.008 and 0.007 
  km$^{-2}$ sr$^{-1}$ yr$^{-1}$ for energy thresholds of 1, 2, 3, 5 and 10 EeV. They  are derived under the conservative choice that the expected background is zero (relevant here only for $E_{0} = 1$~EeV) which makes the limits more robust against hadronic interaction and mass composition assumptions. 
Rescaling the photon flux limits by the measured all-particle spectrum~\cite{AugerSpectrum2} results in  photon fraction limits of 0.14\%, 0.17\%, 0.42\%, 0.86\% and 2.9\% for the same threshold intervals.  \\

The robustness of the results is tested against several sources of systematic uncertainties. Some of them (see table~\ref{tab:syst}) are related to the detector knowledge and the data reconstruction. A contribution of $\pm 6.4$\% applies to the exposure (gray band in Fig.~\ref{fig:expo}) and is obtained as a quadrature sum of the 4\% uncertainty on the ontime~\cite{exposure2010} and the 5\% uncertainties in the FD trigger efficiency after the fiducial distance cut (section~\ref{sect:dataset}). 
The other terms are due to the uncertainties on the energy scale,  \Xm and $S_{b}$. Since these variables are used in the multi-variate analysis, the impact of their systematic uncertainties on the upper limits is evaluated through altering the data by $\pm 1\sigma_{\rm{syst}}$ and applying the BDT to the new data set. Each variable is considered separately even if a correlation is expected between the systematic uncertainties on \Xm and energy scales because of the event reconstruction and the atmospheric contributions. 
A shift by $\Delta$\Xm~=~$\pm 10$ g/cm$^2$~\cite{Xmax} changes the number of selected candidates by $^{+1}_{-2}$ in the first energy interval ($E_0 > 1$~EeV) and leaves unaffected the limits at larger energy thresholds. The same result is obtained when applying a shift of $E_{\gamma}$ by $\Delta E = \pm14$\%~\cite{Verzi2013}. 
The systematic uncertainties on $S_b$ are mostly due to the time synchronization between SD and FD and the possible misalignment of the telescopes which can affect the geometry reconstruction. The latter is periodically tested using lasers and time periods having misaligned mirrors are rejected from the analysis. The SD/FD synchronization is checked using dedicated lasers shots which are observed by the FD and for which a signal is simultaneously sent to an SD station connected to CLF through an optical fiber. Moreover,  
the discrepancy between the core position reconstructed in hybrid and in SD-only modes $-$  having independent systematics on the geometry reconstruction $-$ are compared in data and in simulations. The difference between data and simulations is about +10 m in both easting and northing coordinates and independent of the zenith angle. It translates in a variation of $S_b$ by less than 5\%. 
When applying a shift by $\Delta S_b  = \pm 5\%$ to data, the number of candidates changes by $^{-1}_{+1}$ in the energy range $1-2$ EeV.  The relative change in the upper limits when each of the sources of systematic uncertainty is considered separately is given in the last column of table~\ref{tab:syst}. 
 As an additional test, an altered data set is generated applying a combined shift  (+$\Delta$\Xm, +$\Delta E$, -$\Delta S_{b}$) which would make data more similar to photon events. The number of candidates found in this scenario is 11, 1, 0, 0 and 0 above 1, 2, 3, 5 and 10 EeV. Six of the candidates between $1-2$~EeV were initially at energy below 1~EeV and the candidate with energy above 2~EeV was previously not selected by the BDT cut. 
The maximum range of variation of the upper limits when considering all the experimental systematics  (data and exposure) is shown in Fig.~\ref{fig:limits} as horizontal segments delimiting a dashed-filled box around each energy thresholds. 
Other contributions, related to the assumptions used to train the BDT and select photon-like events, have been considered and for each of them the full analysis, including BDT and selection optimization, data processing and exposure calculation have been performed.  The selected number of candidates and the derived limits are summarized in the table~\ref{tab:models} for each of the tested models. To take into account the lack of knowledge on the hadronic interaction models and the mass composition, the search for photons has been performed using the Epos-LHC model and a proton-iron mix, respectively. 
Moreover, given the large uncertainties on the predicted flux of GZK photons, strongly dependent on the astrophysical scenarios, and for consistency with previous results, a simple power-law assumption with $\Gamma = 2$ is used in the paper as baseline. In the table~\ref{tab:models}, an estimate of the upper limits variation is provided in a range of  values describing possible GZK photon fluxes.


\begin{table}
\begin{center}
  \begin{tabularx}{0.6\textwidth}{c| *{6}{Y}}
     E$_{0}$ [EeV] &  1  &  2 & 3 & 5 & 10 \\ 
     \hline
      \hline
  &\multicolumn{5}{c}{Hadronic model (Epos LHC)}\\
  \cline{2-6}
  N$_{\gamma}$     & 7 & 1 & 0 & 0 &  0\\
  $\Phi^{95\% \rm{C.L.}}$ & 0.043 & 0.015 & 0.008 & 0.008 &  0.008\\
  \hline
  \hline
  &\multicolumn{5}{c}{Mixed composition}\\
   \cline{2-6}
  N$_{\gamma}$     & 2 & 0 & 0 & 0 &  0\\
  $\Phi^{95\% \rm{C.L.}}$ & 0.041 & 0.019 & 0.008 & 0.007&  0.007\\
  \hline
  \hline
  &\multicolumn{5}{c}{Spectral Index $\Gamma = 2.5$} \\ 
   \cline{2-6}
  N$_{\gamma}$     & 6 & 1 & 0 & 0 &  0\\
  $\Phi^{95\% \rm{C.L.}}$ & 0.046 & 0.017 & 0.010 & 0.009 & 0.009\\
\hline
\hline
  &\multicolumn{5}{c}{Spectral Index $\Gamma = 1.5$} \\ 
   \cline{2-6}
  N$_{\gamma}$     & 3 & 0 & 0 & 0 &  0\\
  $\Phi^{95\%\rm{C.L.}}$ & 0.025 & 0.008 & 0.008 & 0.007 &  0.006\\
    \hline
   \hline
  \end{tabularx}
  \end{center}
  \caption{Impact of systematic uncertainties related to the model assumptions on the number of candidates ($N_{\gamma}$) and on the upper limits ($\Phi^{95\%\rm{C.L.}}$, in km$^{-2}$sr$^{-1}$yr$^{-1}$.). For each model, the BDT is trained and is applied to select photon-like events in data and to calculate the exposure. }
    \label{tab:models}
  \end{table}
  
\section{Discussion and conclusions}\label{sect:conclusions}

The upper limits derived in this paper are drawn in Fig.~\ref{fig:limits} compared to other experimental results and to the photon flux predicted for the GZK and the top-down models. In the previous paper~\cite{AugerPhotonHybrid} hybrid events with large \Xm were used to search for photons above 2, 3, 5 and 10~EeV. Eight candidates were found in the first two energy intervals and upper limits were derived on the fraction of photons in the all-particle spectrum. 
The new results lower the upper limits on the photon fraction by a factor 4 at energies above 5 and 10~EeV and up to a factor 25 at $E_{\rm{thr}}= 2$~EeV. This is a consequence of the larger exposure - which equally affects all energy intervals and is responsible for the factor 4 improvement in the two highest energy bins -  and the reduced background contamination which explains the remaining gain at low energies.
The factor 4 increase of the exposure is mostly due to the accumulation of 6 years of data. An additional gain arises from the accurate calculation of the exposure based on time-dependent simulation, avoiding the application of a fiducial cut used in the past to mitigate the dependence of the detector acceptance on mass composition~\cite{AugerPhotonHybrid,AugerPhotonHybrid2007}. 
Moreover, the present analysis based on a BDT and on the combination of SD and FD observables achieves a background contamination of about 10$^{-3}$ ($\sim 4\,\cdot\,10^{-4}$)  for  protons (proton-iron mix), which is at least 10 times lower compared to previous estimations~\cite{AugerPhotonHybrid, AugerPhotonHybridICRC} and has also allowed
extending the analysis down to 1 EeV. 

Some top-down scenarios proposed to explain the origin of trans-GZK cosmic rays (dashed lines) are illustrated though mostly rejected by previous bounds on the photon flux. A recent super-heavy dark matter proposal (SHDM II) developed in the context of an inflationary theory is shown as a long-dashed line. The case of a SHDM particle with mass $M_{\chi}=4.5\times10^{22}$ eV,  life-time $\tau_{\chi}=2.2 \times 10^{22}$ yr and inflaton potential index $\beta = 2$ is only marginally compatible with the limits presented in this work and severely constrained by the limits from the surface detector data~\cite{AugerPhotonSD2015}, in agreement with the interpretation of the Planck results in ~\cite{Planck}. Constraints on the lifetime-and-mass parameter space of SHDM particle can be imposed by current and future limits on the photon flux, as obtained for example in~\cite{ConstrainsSHDM}.  

The achieved sensitivity allows testing photon fractions of about 0.1\% and exploring the region of photon fluxes predicted in some optimistic astrophysical scenarios (GZK proton-I in Fig.~\ref{fig:limits})~\cite{Gelmini}. 
A significant increase of the exposure is required to test more recent proton scenarios~\cite{Kampert} (GZK proton-II in the figure) assuming a maximum acceleration energy of 10$^{21}$~eV and a strong evolution of the source which is only partially constrained by the limits on the neutrino flux above 10~PeV~\cite{Aartsen:2016ngq}.  Under similar astrophysical assumption but with the acceleration of iron primaries at the source, the predicted flux of cosmogenic photons is suppressed by a factor 10. 
Extrapolating the present analysis up to 2025 would reach
 flux limits of a few times 10$^{-3}$ km$^{-2}$ sr$^{-1}$ yr$^{-1}$ at the EeV energies which is at the upper edge of the GZK proton-II expected flux region. A factor 10 larger statistics can be gained with a future  SD-based analysis above about 10$^{18.5}$~eV 
using new SD triggers that have been installed in all array stations and that are designed to enhance photon and neutrino detection efficiencies~\cite{NIM2015, UHECR2014}. 
A deployment of a 4 m$^2$ scintillator on top of each SD is foreseen as a part of the AugerPrime upgrade of the Observatory to determine the muon content of the air-showers at the ground which may provide further information to distinguish between photon- and hadron-induced showers~\cite{AugerPrime}.


\section*{Acknowledgments}

\begin{sloppypar}
The successful installation, commissioning, and operation of the Pierre Auger Observatory would not have been possible without the strong commitment and effort from the technical and administrative staff in Malarg\"ue. We are very grateful to the following agencies and organizations for financial support:
\end{sloppypar}

\begin{sloppypar}
Argentina -- Comisi\'on Nacional de Energ\'\i{}a At\'omica; Agencia Nacional de Promoci\'on Cient\'\i{}fica y Tecnol\'ogica (ANPCyT); Consejo Nacional de Investigaciones Cient\'\i{}ficas y T\'ecnicas (CONICET); Gobierno de la Provincia de Mendoza; Municipalidad de Malarg\"ue; NDM Holdings and Valle Las Le\~nas; in gratitude for their continuing cooperation over land access; Australia -- the Australian Research Council; Brazil -- Conselho Nacional de Desenvolvimento Cient\'\i{}fico e Tecnol\'ogico (CNPq); Financiadora de Estudos e Projetos (FINEP); Funda\c{c}\~ao de Amparo \`a Pesquisa do Estado de Rio de Janeiro (FAPERJ); S\~ao Paulo Research Foundation (FAPESP) Grants No.\ 2010/07359-6 and No.\ 1999/05404-3; Minist\'erio de Ci\^encia e Tecnologia (MCT); Czech Republic -- Grant No.\ MSMT CR LG15014, LO1305 and LM2015038 and the Czech Science Foundation Grant No.\ 14-17501S; France -- Centre de Calcul IN2P3/CNRS; Centre National de la Recherche Scientifique (CNRS); Conseil R\'egional Ile-de-France; D\'epartement Physique Nucl\'eaire et Corpusculaire (PNC-IN2P3/CNRS); D\'epartement Sciences de l'Univers (SDU-INSU/CNRS); Institut Lagrange de Paris (ILP) Grant No.\ LABEX ANR-10-LABX-63 within the Investissements d'Avenir Programme Grant No.\ ANR-11-IDEX-0004-02; Germany -- Bundesministerium f\"ur Bildung und Forschung (BMBF); Deutsche Forschungsgemeinschaft (DFG); Finanzministerium Baden-W\"urttemberg; Helmholtz Alliance for Astroparticle Physics (HAP); Helmholtz-Gemeinschaft Deutscher Forschungszentren (HGF); Ministerium f\"ur Innovation, Wissenschaft und Forschung des Landes Nordrhein-Westfalen; Ministerium f\"ur Wissenschaft, Forschung und Kunst des Landes Baden-W\"urttemberg; Italy -- Istituto Nazionale di Fisica Nucleare (INFN); Istituto Nazionale di Astrofisica (INAF); Ministero dell'Istruzione, dell'Universit\'a e della Ricerca (MIUR); CETEMPS Center of Excellence; Ministero degli Affari Esteri (MAE); Mexico -- Consejo Nacional de Ciencia y Tecnolog\'\i{}a (CONACYT) No.\ 167733; Universidad Nacional Aut\'onoma de M\'exico (UNAM); PAPIIT DGAPA-UNAM; The Netherlands -- Ministerie van Onderwijs, Cultuur en Wetenschap; Nederlandse Organisatie voor Wetenschappelijk Onderzoek (NWO); Stichting voor Fundamenteel Onderzoek der Materie (FOM); Poland -- National Centre for Research and Development, Grants No.\ ERA-NET-ASPERA/01/11 and No.\ ERA-NET-ASPERA/02/11; National Science Centre, Grants No.\ 2013/08/M/ST9/00322, No.\ 2013/08/M/ST9/00728 and No.\ HARMONIA 5 -- 2013/10/M/ST9/00062; Portugal -- Portuguese national funds and FEDER funds within Programa Operacional Factores de Competitividade through Funda\c{c}\~ao para a Ci\^encia e a Tecnologia (COMPETE); Romania -- Romanian Authority for Scientific Research ANCS; CNDI-UEFISCDI partnership projects Grants No.\ 20/2012 and No.194/2012 and PN 16 42 01 02; Slovenia -- Slovenian Research Agency; Spain -- Comunidad de Madrid; Fondo Europeo de Desarrollo Regional (FEDER) funds; Ministerio de Econom\'\i{}a y Competitividad; Xunta de Galicia; European Community 7th Framework Program Grant No.\ FP7-PEOPLE-2012-IEF-328826; USA -- Department of Energy, Contracts No.\ DE-AC02-07CH11359, No.\ DE-FR02-04ER41300, No.\ DE-FG02-99ER41107 and No.\ DE-SC0011689; National Science Foundation, Grant No.\ 0450696; The Grainger Foundation; Marie Curie-IRSES/EPLANET; European Particle Physics Latin American Network; European Union 7th Framework Program, Grant No.\ PIRSES-2009-GA-246806; and UNESCO.
\end{sloppypar}

\end{document}